\documentclass[reprint,prb,aps,twocolumn,superscriptaddress,  10pt]{revtex4-2}
\usepackage{amsmath}
\usepackage{amssymb}
\usepackage[english]{babel}
\usepackage{graphicx}
\usepackage{cancel} 
\usepackage{soul}
\usepackage{xcolor}
\usepackage{caption}
\usepackage{ragged2e}
\DeclareCaptionJustification{justified}{\justifying}
\AtBeginDocument{\captionsetup[figure]{labelsep=period, labelfont=bf, justification=justified,
    singlelinecheck=false, width=\columnwidth}}

\newcommand{\ket}[1]{|#1\rangle}

\def \Technion{The Physics Department and the Solid State Institute,
Technion--Israel Institute of Technology, 32000 Haifa, Israel}

\begin{document}

\title{Restoring polarization entanglement from solid-state photon sources by time-dependent photonic control}

\author{Ismail Nassar}
\affiliation{\Technion}
\author{Dan Cogan}
\affiliation{\Technion}
\author{Ido Schwartz}
\affiliation{\Technion}

\begin{abstract}
Quantum states of light are central resources for quantum communication, networking, and photonic information processing. In many quantum emitters, coherent internal dynamics arising from intrinsic or field-induced level splittings imprint a deterministic, time-dependent phase on the emitted light. When emission times are stochastic and detector timing resolution is finite, this phase evolution becomes effectively unresolved, suppressing observable entanglement.

Here, we demonstrate a photonic-compensation protocol that removes this emitter-induced phase evolution directly in the photonic domain. Rather than modifying the emitter, we apply synchronized, time-dependent coherent operations to the emitted photons that reverse the accumulated phase independently of the emission time. Using exciton fine-structure splitting in a semiconductor quantum dot as a model system, we implement dynamic phase modulation and perform time-resolved two-photon polarization tomography. We show that this restores a stationary two-photon polarization state and recovers polarization entanglement without temporal post-selection and independently of detector timing resolution.

Our approach provides a scalable route to robust solid-state entangled-photon sources and, more broadly, establishes a strategy for removing the imprint of coherent emitter dynamics on photonic entanglement in integrated platforms.

\end{abstract}

\maketitle

\section{Introduction}
Quantum states of light are central resources for quantum technologies, spanning single-photon states, entangled photon pairs, and multipartite photonic entanglement. They underpin secure communication~\cite{ekert1991}, quantum-state teleportation~\cite{Bennett1993}, and photonic quantum computation~\cite{Knill2001,kok2007linear}. A central challenge is to realize sources that are simultaneously deterministic, high-fidelity, and scalable in solid-state platforms.

A widespread obstacle in many quantum emitters is that coherent internal evolution during photon emission imprints a deterministic, time-dependent phase on the emitted light. Level splittings and related dynamics map onto the emitted photonic state as a time-dependent relative phase between radiative pathways~\cite{Wilk2007,Togan2010,Christle2017,Kolesov2013,Akopian2006}. Although this phase evolution is deterministic, it is referenced to stochastic emission times. When experiments average over these emission delays, and when detector timing resolution is finite, the phase becomes effectively unresolved. As a result, the reconstructed photonic state appears mixed and the observable entanglement is suppressed.

Here, we pursue a conceptually new route to overcome this obstacle: rather than engineering the emitter to eliminate the coherent dynamics, we reverse their phase imprint after emission using synchronized, time-dependent photonic operations~\cite{Varo2022,Fognini18}. In doing so, we undo the deterministic phase evolution mapped onto the photons and convert a non-stationary photonic quantum state into a stationary one on an event-by-event basis, without temporal post-selection.

Semiconductor quantum dots (QDs) provide a paradigmatic setting for this problem and an attractive platform for on-demand entangled-photon generation via the biexciton-exciton radiative cascade~\cite{Akopian2006,stevenson2006,Dousse2010,Muller2014,Schmidgall2015}. Ideally, this cascade emits a maximally entangled polarization Bell state. In practice, exciton fine-structure splitting (FSS)~\cite{Gammon1996,Winik2017} lifts the degeneracy of the intermediate exciton states and induces spin precession, imprinting a well-defined phase evolution on the two decay paths. When averaged over the random biexciton-to-exciton delay, this phase encodes which-path information and washes out the observable polarization entanglement.

Considerable effort has therefore focused on suppressing the FSS below the radiative linewidth using strain tuning~\cite{Zhang2015}, applied electric and magnetic fields~\cite{Bennett2010,Stevenson2006_mag}, and engineered cavity architectures~\cite{Lodahl2015}. While powerful, these approaches require careful device optimization, typically operate over limited ranges, and are difficult to implement uniformly across large ensembles. Residual coherent dynamics often remain and become detrimental whenever detector timing cannot fully resolve the evolution.

In this work, we implement photonic compensation for a single QD driven by deterministic two-photon excitation of the biexciton~\cite{Muller2014,Winik2017}. We apply dynamic phase modulation (DPM) synchronized to the excitation clock and matched to the exciton precession frequency set by the FSS. Using time-resolved, full two-photon polarization-state tomography~\cite{Kwiat2001,Akopian2006,Winik2017}, we show that DPM restores a stationary two-photon polarization state and recovers high-fidelity polarization entanglement without temporal filtering and independently of detector timing resolution. More broadly, this approach applies whenever coherent emitter evolution maps onto the emitted photons as a deterministic phase referenced to a clock, providing a scalable route to robust entangled-photon sources in solid-state and integrated photonic platforms.

\section{Results}

\begin{figure}[h!]
\centering
\includegraphics[width=\columnwidth]{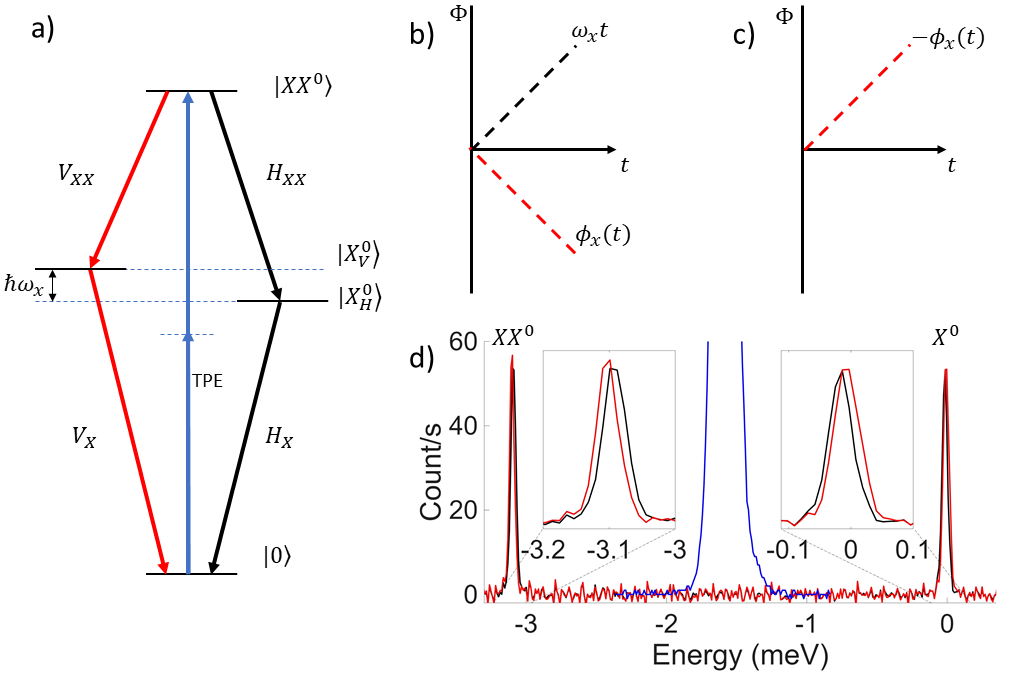}
\caption{
Concept of dynamic phase modulation.
(a) Energy-level diagram of the biexciton–exciton cascade showing fine-structure splitting (FSS) and polarization-dependent decay paths.
(b) Temporal phase dependence of the exciton (black) and the emitted photon compensating DPM (red).
(c) Phase modulation applied to the biexciton photon to remove the time dependence from the $XX^0$–$X^0$ cascade.
(d) Emission spectrum under two-photon excitation (blue). Black and red curves correspond to horizontal (H) and vertical (V) polarization projections, respectively. Insets show zooms of the $XX^0$ and $X^0$ emission lines.
}
\label{energy_diagram_spectrum}
\end{figure}

Before describing the experiment, we outline the operating principle of dynamic phase modulation (DPM; see Supplementary Note~1 for a full derivation). In the biexciton-exciton cascade of a semiconductor quantum dot (Fig.~\ref{energy_diagram_spectrum}a), coherent emitter dynamics, here the exciton fine-structure splitting $\Delta_{\mathrm{FSS}}$, map onto the emitted photons as a deterministic, time-dependent relative phase between the two radiative pathways. Because the biexciton-to-exciton delay varies stochastically from event to event, and because detector timing resolution is finite, experiments effectively average over this phase evolution, washing out polarization coherences and suppressing the observable entanglement. The resulting two-photon state can be written as~\cite{Winik2017}
\begin{equation}
\label{Eq. phase evolution}
\ket{\psi_{2ph}(t)}=\frac{1}{\sqrt{2}}\big(\ket{H_{XX}H_X}+e^{-i\omega_X(t_X-t_{XX})}\ket{V_{XX}V_X}\big),
\end{equation}
where $\omega_X=\Delta_{\mathrm{FSS}}/\hbar$ is the exciton precession frequency and $t_{XX}$ ($t_X$) denotes the biexciton (exciton) emission time.

Rather than suppressing the fine-structure splitting at the emitter, we compensate its imprint directly in the photonic domain. Specifically, an electro-optic modulator (EOM) implements a phase $\phi(t)$ between the $H$ and $V$ polarization components, synchronized to the excitation clock. Choosing a linear phase ramp with slope $\dot{\phi}=\omega_X$ cancels the exciton-induced phase evolution in Eq.~\ref{Eq. phase evolution} on an event-by-event basis, yielding a time-independent two-photon state,
\begin{equation}
\ket{\psi_{2ph}}=\frac{1}{\sqrt{2}}\big(\ket{H_{XX}H_X}+\ket{V_{XX}V_X}\big),
\end{equation}
i.e., the Bell state $\ket{\Phi^{+}}$. This photonic-compensation concept, proposed theoretically in Refs.~\cite{Varo2022,Fognini18}, restores polarization entanglement without modifying the quantum dot. In the following, we experimentally implement this protocol and verify the recovery of a stationary polarization-entangled state (see Supplementary Note~1 for the DPM formalism).

Figure~\ref{energy_diagram_spectrum}d shows polarization-resolved photoluminescence (PL) from a single quantum dot under resonant two-photon excitation of the biexciton. The excitation laser is tuned to the two-photon resonance midway between the $XX^0$ and $X^0$ transitions, enabling deterministic preparation of the biexciton state. The measured PL spectrum exhibits the expected linearly polarized $XX^0$ and $X^0$ doublets, separated by a fine-structure splitting of $\Delta_{\mathrm{FSS}} = 8.80 \pm 0.04\,\mu\text{eV}$. This splitting sets the exciton precession frequency $\omega_X=\Delta_{\mathrm{FSS}}/\hbar$ and corresponds to a precession period of $T_p = h/\Delta_{\mathrm{FSS}} = 470 \pm 2\,\text{ps}$, as illustrated in Figs.~\ref{energy_diagram_spectrum}b--c.

\begin{figure*}
\centering
\includegraphics[width=\textwidth]{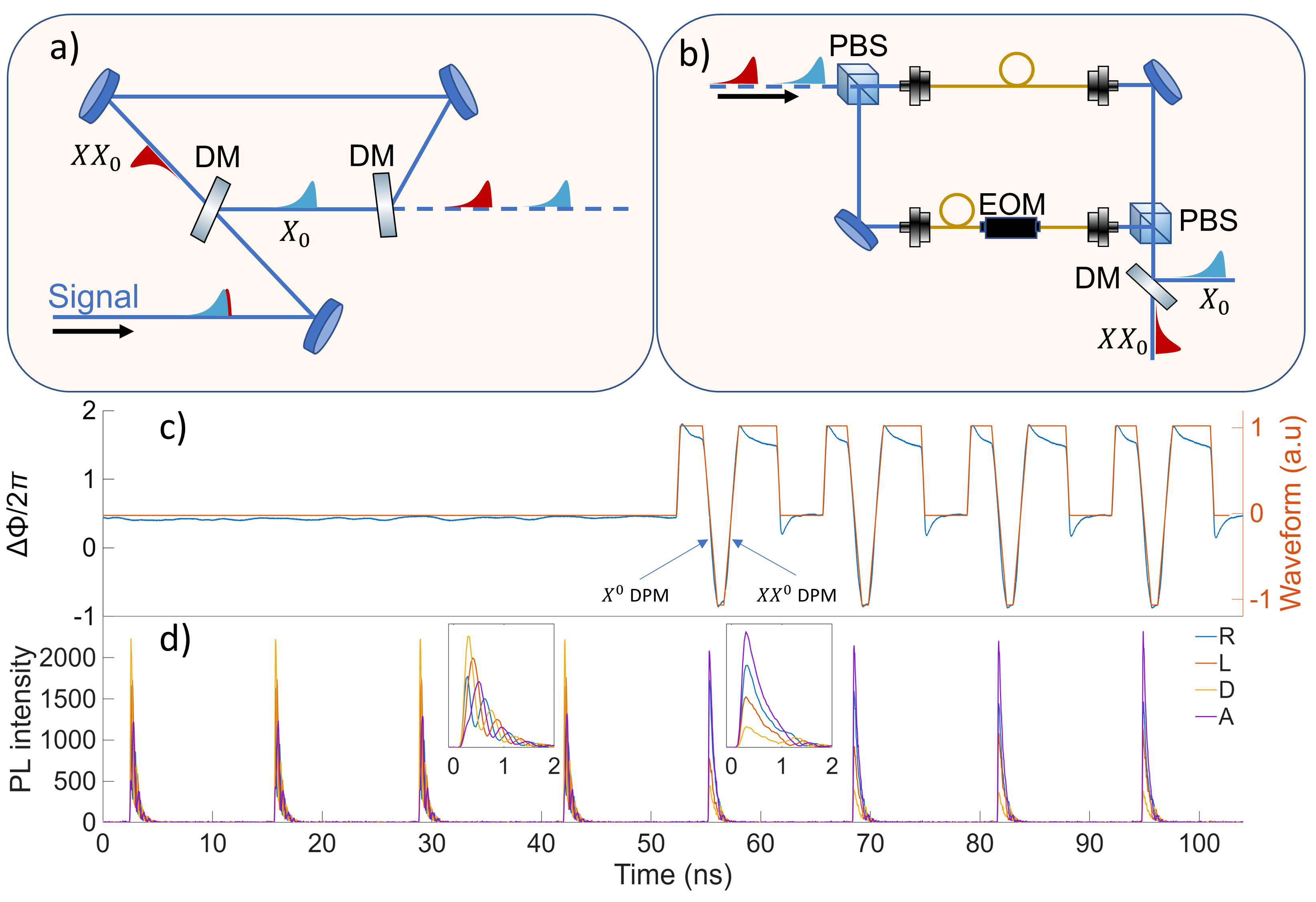}
\captionsetup{width=\textwidth}
\caption{
Dynamic phase modulation of the cascade photons.
    (a) Temporal separation of the $XX^0$ and $X^0$ photons by $\Delta t=1.9\,\text{ns}$ using a pair of dichroic mirrors (DMs). The exciton (blue) and biexciton (red) paths are indicated by exponential decay profiles.
    (b) Schematic of the Mach–Zehnder interferometer with an integrated phase modulator in one arm. A polarizing beam splitter (PBS) separates $H$ and $V$ polarization components, which are recombined at a second beam splitter. After the interferometer, a dichroic mirror (DM) spatially separates the photons.
    (c) Temporal waveform applied to the phase modulator (red) and a histogram of the induced phase (blue) measured using a $D$-polarized CW laser tuned to the exciton wavelength.
    (d) Histogram of time-resolved PL, showing eight consecutive emission pulses projected into four polarizations (R, L, D, A). Insets show the sum over the first four pulses (left, without DPM) and the last four pulses (right, with DPM).
}
\label{experimental_schematics}
\end{figure*}


We next measure two-photon polarization correlations under deterministic biexciton preparation using a $\pi$-area two-photon excitation (TPE) pulse applied to an optically depleted QD~\cite{Schmidgall2015}. Following excitation, the biexciton-exciton cascade emits the $XX^0$ and $X^0$ photons sequentially. In the presence of exciton FSS, the stochastic biexciton-to-exciton delay maps onto a fluctuating relative phase between the two decay paths, and finite detector timing resolution further averages over this evolution. As a result, polarization coherences are strongly reduced and the observable entanglement can be severely degraded, or even vanish, without temporal post-selection~\cite{Winik2017}.

To implement this compensation in practice, the applied phase must remove the stochastic relative phase $\omega_X(t_X-t_{XX})$ in Eq.~\ref{Eq. phase evolution}. This is achieved using two complementary phase ramps referenced to the same excitation clock: a ramp that cancels the $-\omega_X t_X$ term during the $X^0$ wavepacket, and a ramp that cancels the $+\omega_X t_{XX}$ term during the delayed $XX^0$ wavepacket. We therefore spectrally separate the $XX^0$ and $X^0$ photons and route the biexciton photon through a delay line that imposes a well-defined differential delay of $\Delta t = 1.9\,\text{ns}$. The two photons are then recombined into a common spatial mode and directed into a fiber-based Mach-Zehnder interferometer (Fig.~\ref{experimental_schematics}a). Owing to the imposed delay, the $XX^0$ and $X^0$ photons traverse the interferometer at distinct times. A polarizing beam splitter separates the $H$ and $V$ components into different arms, and an electro-optic modulator (EOM) placed in the $V$ arm applies the programmed dynamic phase modulation synchronized to the excitation pulses (Fig.~\ref{experimental_schematics}b). After the interferometer, a narrow band-pass filter transmits the biexciton photon and reflects the exciton photon, enabling independent polarization analysis using liquid-crystal variable retarders (LCVRs) and a polarizing beam splitter. Photon arrival times are recorded using superconducting nanowire single-photon detectors and a time-tagging module (see Supplementary Fig.~S2). Active stabilization of the Mach-Zehnder interferometer maintains phase stability throughout the long integrations required for tomography.

The temporal waveform driving the EOM is shown in Fig.~\ref{experimental_schematics}c. Each modulation cycle contains eight excitation pulses: the first four serve as an unmodulated reference, while the subsequent four implement the programmed DPM. Prior to the arrival of the $X^0$ photon, and throughout its radiative decay, the EOM voltage is ramped down linearly at a slope matched to $\omega_X$, applying a phase $+\omega_X t$ across the $X^0$ wavepacket, thereby cancelling the factor $e^{-i\omega_X t_X}$ in Eq.~\ref{Eq. phase evolution}. Before the delayed $XX^0$ photon arrives, the voltage is ramped up with the opposite slope, applying a phase $-\omega_X t$ that cancels the factor $e^{+i\omega_X t_{XX}}$. The combined waveform therefore removes the full stochastic phase $\omega_X(t_X-t_{XX})$, stabilizing the two-photon polarization state for every emission event. 

The induced phase evolution was verified by transmitting a continuous-wave $D$-polarized laser through the MZI and projecting the output onto the $R$, $L$, $D$, and $A$ bases (Fig.~\ref{experimental_schematics}c, blue histogram). The inferred phase evolution faithfully tracks the programmed waveform, aside from a constant offset arising from the interferometer bias point. 
We further verified the action of the modulation using p-shell resonant excitation of the $X^{0}$ transition~\cite{Schwartz2015prb}. Time-resolved, polarization-sensitive photoluminescence shows clear oscillations in the exciton emission under FSS, which are removed when the negative-slope portion of the waveform is applied (Fig.~\ref{experimental_schematics}d). Thus, DPM dynamically suppresses the exciton precession and yields a time-independent exciton polarization during its radiative lifetime.

\begin{figure}[h!]
\centering
\includegraphics[width=\columnwidth]{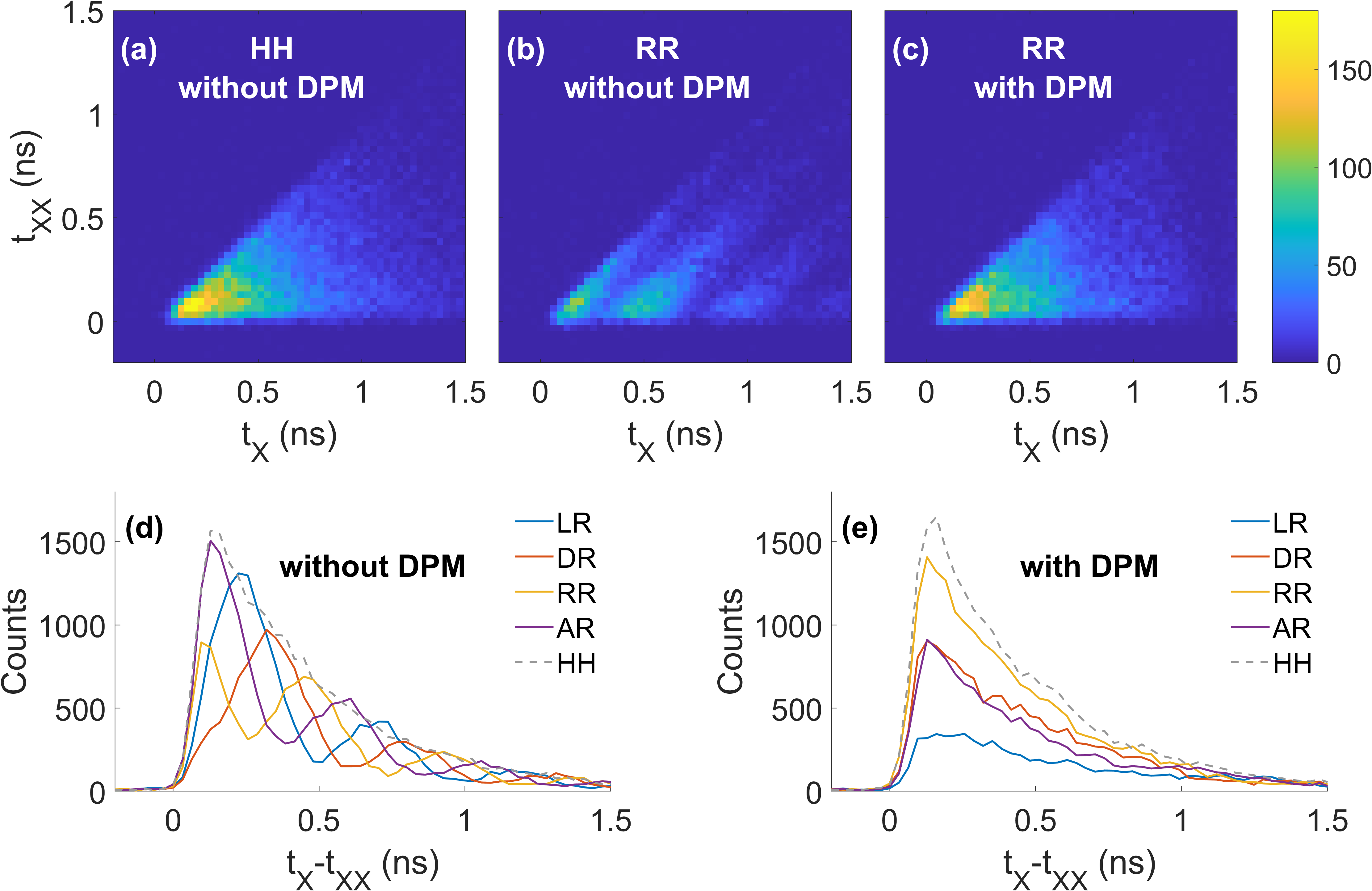}
\caption{
Coincidence measurements of biexciton–exciton photon pairs.
    (a–c) Measured $XX^0$–$X^0$ coincidence maps as a function of detection times for representative polarization bases: collinear (HH) without DPM (a), co-circular (RR) without DPM (b), and co-circular (RR) with DPM (c). The capital letters denote, respectively, the biexciton and exciton polarization projections. The color scale indicates the number of coincidences per $16\,\text{ps}\times16\,\text{ps}$ bin.
    (d,e) Coincidence rate as a function of the arrival-time difference between $X^0$ and $XX^0$ photons ($t_{X}-t_{XX}$) for various polarization bases, without DPM (d) and with DPM (e).
}
\label{2D_XX0_X0_heatmaps}
\end{figure}

Having established dynamic suppression of the exciton phase, we next examine the polarization correlations between the $XX^{0}$ and $X^{0}$ photons under deterministic biexciton generation via TPE. Figures~\ref{2D_XX0_X0_heatmaps}a–c show coincidence maps of the photon detection times for representative two-photon polarization projections. In the collinear $HH$ basis (Fig.~\ref{2D_XX0_X0_heatmaps}a), no oscillations are observed, consistent with projecting the exciton onto the eigenstate $\ket{X^{0}_{\mathrm{H}}}$~\cite{Winik2017}. By contrast, in the co-circular $RR$ basis (Fig.~\ref{2D_XX0_X0_heatmaps}b), the coincidences exhibit pronounced oscillations arising from exciton spin precession. When DPM is applied (Fig.~\ref{2D_XX0_X0_heatmaps}c), these oscillations disappear entirely, demonstrating that the joint two-photon phase is time independent.

This behaviour is quantified in Figs.~\ref{2D_XX0_X0_heatmaps}d–e, where diagonal integration of the coincidence maps yields the coincidence rate as a function of the arrival-time difference $t_{X}-t_{XX}$. Without DPM, clear oscillations appear in all non-eigenstate bases and exhibit the expected $\pi/2$ phase shifts across the sequence $LR \rightarrow DR \rightarrow RR \rightarrow AR$ (Fig.~\ref{2D_XX0_X0_heatmaps}d). When DPM is applied, these oscillations are fully suppressed in every basis (Fig.~\ref{2D_XX0_X0_heatmaps}e), consistent with complete cancellation of the phase accumulated during the cascade (see all projections in Supplementary Fig.~S4). The oscillations constitute the time-domain signature of FSS-induced exciton precession, which encodes which-path information and suppresses polarization entanglement. Their disappearance under DPM confirms that this which-path information is erased and that entanglement is restored~\cite{Winik2017}.

\begin{figure}[b!]
\centering
\includegraphics[width=\columnwidth]{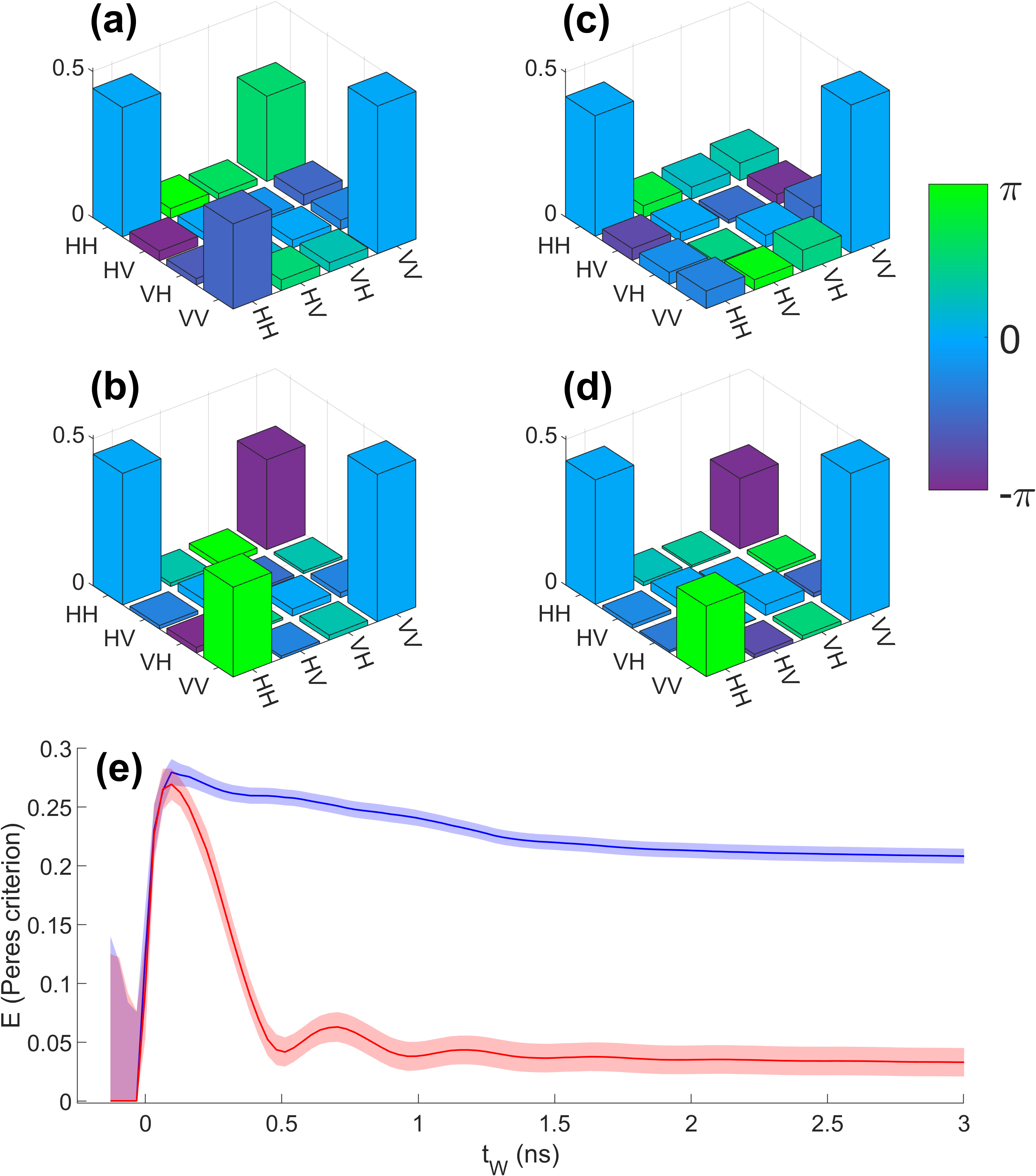}
\caption{
    Two-photon density matrices and entanglement negativity.
    (a,b) Measured two-photon polarization density matrices for a narrow integration window $t_W = 0.096\,\text{ns}$ without DPM (a) and with DPM (b).
    (c,d) Density matrices for a wide integration window $t_W = 3\,\text{ns}$ without DPM (c) and with DPM (d). The color bar represents the phase of the density matrix elements.
    (e) Negativity $\mathcal{N}$ of the two-photon density matrix as a function of the integration window $t_W$, without DPM (red) and with DPM (blue). The shaded regions correspond to experimental uncertainty of three standard deviations.
}
\label{Density_matrices_Negativity}
\end{figure}

To quantify the effect of DPM on the joint state, we reconstructed the two-photon density matrix~\cite{Kwiat2001} from 36 polarization projections for a range of temporal detection windows $t_W$, chosen with equal widths for the $XX^0$ and $X^0$ photons. For a narrow window of $t_W = 0.096\,\text{ns}$, the reconstructed matrices with and without DPM (Figs.~\ref{Density_matrices_Negativity}a–b) are nearly identical, apart from a global phase offset originating from the precise photon arrival time relative to the applied modulation. These matrices provide a baseline for comparison with wider temporal windows, where phase averaging becomes significant.

For a much wider integration window of $t_W = 3\,\text{ns}$, the impact of DPM becomes striking. Without modulation (Fig.~\ref{Density_matrices_Negativity}c), phase averaging over many decay events suppresses the off-diagonal coherences, yielding an almost classical mixture. With DPM applied (Fig.~\ref{Density_matrices_Negativity}d), these coherences persist with well-defined phases across the entire window, demonstrating that the polarization entanglement is preserved throughout the full radiative cascade.

We quantify the two-photon entanglement using the negativity $\mathcal{N}$, a basis-independent measure that reliably detects non-separability even for mixed states~\cite{Peres1996,Vidal2002}. The dependence of $\mathcal{N}$ as a function of the temporal detection window $t_W$ is shown in Fig.~\ref{Density_matrices_Negativity}(e). For narrow windows, we obtain $\mathcal{N}\approx 0.28$, below the Bell-state value of $0.5$ owing to experimental imperfections (see Supplementary Methods). Without DPM, $\mathcal{N}$ decreases rapidly as $t_W$ increases and exhibits oscillations at the exciton precession period~\cite{Winik2017}. With DPM, however, the negativity remains high and nearly constant across nanosecond-scale windows, reaching $\sim\,0.21$ for the widest interval measured. The pronounced separation between the two curves demonstrates that DPM preserves entanglement irrespective of the fine-structure splitting or detector timing resolution.

The reduction in the measured negativity arises entirely from technical rather than fundamental limitations. 
(i) The applied phase does not track the programmed waveform perfectly, leaving small residual errors. 
(ii) The EOM offers a limited dynamic range (about $4\pi$, equivalent to two exciton rotations), and the photon arrival was delayed by $\sim 300\,\text{ps}$ to avoid nonlinear transients, thereby further narrowing the effective compensation window. 
(iii) Imperfect extinction of the $\ket{H}$ and $\ket{V}$ components in the interferometer arms produces a slight tilt in the rotation axis. 
(iv) Slow interferometer drifts accumulate over long integrations. 
In combination, these effects account for the deviation of the measured negativity from the Bell-state value. All are technical in nature and can be mitigated using modulators with larger phase range, higher-extinction optics, and active waveform and phase optimization.


\section{Discussion}

We experimentally implement a photonic-compensation protocol~\cite{Varo2022,Fognini18} that restores polarization entanglement in the biexciton-exciton cascade by reversing the deterministic phase evolution imprinted by the emitter. Rather than modifying the quantum dot to suppress fine-structure splitting, we cancel the accumulated time-dependent phase using clock-referenced dynamic phase modulation in the photonic domain. This converts a non-stationary two-photon state, whose coherence is obscured by stochastic emission times and finite detector timing resolution, into a stationary two-photon polarization state on an event-by-event basis, without temporal post-selection and independently of detector timing resolution. The experiment therefore shows that the apparent loss of entanglement originates not from irreversible decoherence, but from an unresolved yet deterministic unitary evolution that can be inverted.

Because the exciton precesses at a well-defined frequency set by the fine-structure splitting, the phase accumulated between the two decay pathways is deterministic even though the emission times themselves are stochastic. By referencing linear phase ramps to the excitation clock, we remove this event-dependent phase for every cascade. Entanglement recovery therefore does not rely on resolving the exciton dynamics in detection~\cite{Winik2017}, but on actively inverting the unitary evolution that maps the emitter’s coherent dynamics onto the emitted photons, as evidenced by the disappearance of time-dependent polarization oscillations and the persistence of entanglement across wide temporal windows. This reframes coherent level splittings from a materials limitation into a controllable photonic operation.

The reconstructed two-photon density matrices directly quantify this restoration of coherence. For narrow temporal windows, the measured negativity reaches $\mathcal{N}\approx 0.28$, below the Bell-state value of $0.5$ owing to experimental imperfections. As the integration window increases, the contrast between the uncompensated and compensated cases becomes pronounced: without DPM, phase averaging suppresses the off-diagonal coherences, the negativity decreases rapidly and exhibits oscillations at the exciton precession period, whereas with DPM the negativity remains high and approximately constant across nanosecond-scale windows. The persistence of entanglement over wide temporal windows shows that polarization coherence is maintained independently of fine-structure splitting and detector timing resolution.

The remaining deviation from the Bell-state limit arises from technical rather than fundamental constraints. These include the finite phase range of the electro-optic modulator, residual mismatch between the programmed and realized phase waveform, non-ideal polarization extinction in the interferometer, and slow phase drifts during long integrations. The limited modulation range and the temporal offset introduced to avoid nonlinear transients further restrict the effective compensation window. Together, these engineering limitations account for the observed reduction in negativity and do not reflect an intrinsic bound of the protocol. Improvements in modulation bandwidth and phase range, optical extinction, and interferometric stabilization should allow negativity values closer to the Bell-state limit.

Beyond the specific case of fine-structure splitting in quantum dots, this work establishes a general strategy for treating coherent emitter dynamics as reversible unitary processes rather than intrinsic sources of entanglement loss. Whenever internal evolution maps deterministically onto emitted photons as a time-dependent phase referenced to a stable clock, readily available under pulsed excitation, synchronized photonic operations can in principle remove this imprint and recover a stationary entangled state. Integration of such dynamic compensation with low-loss interferometers and high-speed phase modulators on chip~\cite{Silverstone2013,Wang2019,Moody2022} could therefore provide a scalable route toward robust solid-state entangled-photon sources for quantum communication and networked photonic architectures. This approach offers a route to scalable sources in regimes where uniform suppression of emitter splittings across many devices is impractical.

\section*{Methods}

\subsection*{Device}
The sample consisted of strain-induced InGaAs quantum dots embedded in a 285-nm intrinsic GaAs layer grown on a [001] GaAs substrate by molecular beam epitaxy and incorporated in a one-wavelength planar microcavity formed by AlAs/GaAs distributed Bragg reflectors (25 pairs below and 11 above). A schematic of the structure is shown in Supplementary Fig.~S1.

The investigated quantum dot exhibited biexciton and exciton lifetimes of $\tau_{XX}=211 \pm 5$ ps and $\tau_X=405\pm3$ ps, respectively, and a fine-structure splitting of $\Delta_{\mathrm{FSS}}=8.80\pm0.04~\mu$eV, corresponding to an exciton precession period of $T_p=470\pm2$ ps.

\subsection*{Experimental setup}

The sample was maintained at $5\,\text{K}$ in a closed-cycle cryostat. Excitation and collection were performed through a 60$\times$ objective (NA = 0.85). Resonant two-photon excitation of the biexciton was achieved using $12\,\text{ps}$ pulses from a synchronously pumped dye laser operating at $76\,\text{MHz}$ and tuned to $967.37\,\text{nm}$. A weak continuous-wave laser at $939.27\,\text{nm}$ was used to depopulate long-lived dark excitons~\cite{Schmidgall2015, Winik2017}.

The emitted $XX^0$ and $X^0$ photons were spectrally separated and recombined after introducing a controlled temporal delay $\Delta t=1.9\,\text{ns}$ between them. Dynamic phase modulation was implemented in a fiber-based Mach–Zehnder interferometer using a polarization-selective electro-optic phase modulator driven by an arbitrary waveform generator synchronized to the excitation pulses. Linear phase ramps matched to the exciton precession frequency $\omega_X=\Delta_\mathrm{FSS}/\hbar$ were applied. A full schematic of the setup is shown in Supplementary Fig.~S2.

Long-term interferometer stability was ensured by active phase stabilization using a continuous-wave reference laser. The stabilization scheme and performance characterization are provided in Supplementary Methods (Interferometer stability) and Supplementary Fig.~S3.

Polarization analysis was performed using liquid-crystal variable retarders and polarizing beam splitters. Photon arrival times were recorded with superconducting nanowire single-photon detectors and time-correlated single-photon counting electronics.

\subsection*{Density matrix tomography}

Two-photon polarization-state tomography was performed using 36 polarization projection settings spanning the $\{H,V,D,A,R,L\}$ bases for both photons. Coincidence histograms were accumulated as a function of the photon arrival-time difference and integrated over selected temporal windows. Physical two-photon density matrices were reconstructed using a maximum-likelihood algorithm assuming Poissonian counting statistics.

Statistical uncertainties were propagated through the reconstruction procedure, and systematic uncertainties associated with polarization-setting accuracy were incorporated using a $\pm3^\circ$ tolerance. Entanglement was quantified using the negativity $\mathcal{N}$ computed via the partial-transpose criterion. Full coincidence maps for all projection settings are shown in Supplementary Fig.~S4.

\section*{Data availability}
The data that support the findings of this study are available on reasonable request.

\bibliographystyle{naturemag}
\bibliography{refs}


\section*{Author Contributions} I. N. prepared the experimental setup, performed the experiment, and analyzed the data with input from D. C and I. S.. I. N. and I. S. wrote the manuscript.
I. S. supervised the project.\\

\section*{Competing interests}
The authors declare no competing financial interests.\\

{\bf Correspondence and requests for materials}
should be addressed to I. S. (idosch@technion.ac.il).

\end{document}


\begin{center}
{\LARGE \textbf{Supplementary Information}}\\[1.5em]
{\large \textbf{Restoring polarization entanglement from solid-state photon sources by time-dependent photonic control}}\\[1em]
Ismail Nassar, Dan Cogan, and Ido Schwartz\\
\textit{The Physics Department and the Solid State Institute,}\\
\textit{Technion--Israel Institute of Technology, 32000 Haifa, Israel}\\[1em]
\textbf{Correspondence:} \href{mailto:idosch@technion.ac.il}{idosch@technion.ac.il}
\end{center}

\vspace{1cm}

\section*{Supplementary Methods}
\subsection*{Device}
\begin{figure}[h!]
\centering
\includegraphics[width=0.5\textwidth]{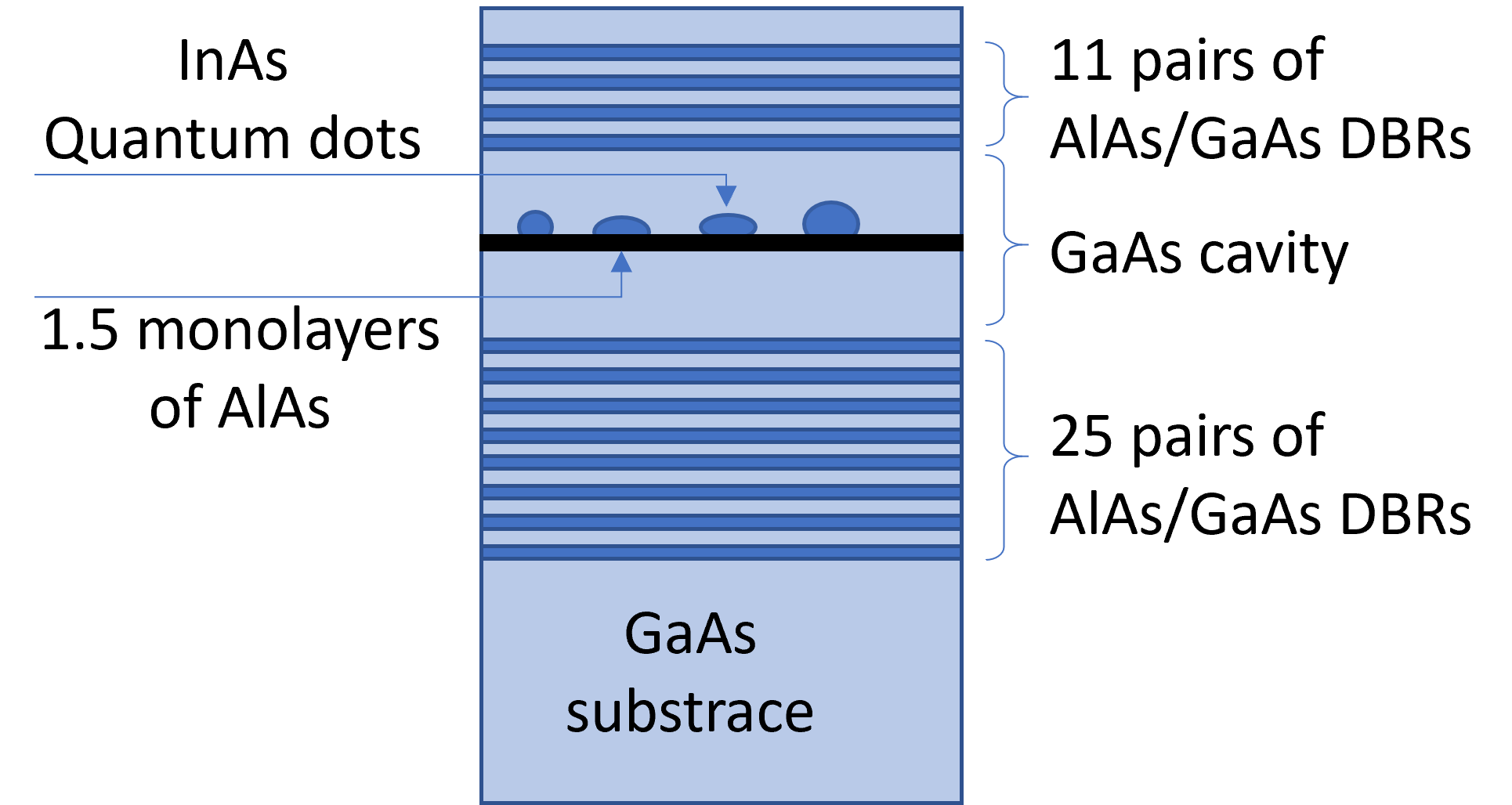}
\caption{
Schematic of the Sample Structure. strain-induced InGaAs quantum dots (QDs) embedded at the center of a 285-nm intrinsic GaAs layer grown on a [001]-oriented GaAs substrate. A one-wavelength ($\lambda$) planar microcavity is formed by distributed Bragg reflectors (DBRs) with 25 AlAs/GaAs quarter-wave pairs below and 11 above, enabling efficient collection of ground-state emission from single QDs resonant with the cavity mode.
}
\label{Sample_structure}
\end{figure}

The sample consisted of strain-induced InGaAs quantum dots (QDs) embedded at the center of a 285-nm intrinsic GaAs layer grown on a [001] GaAs substrate by molecular beam epitaxy. A one-wavelength ($\lambda$) planar cavity was formed by two AlAs/GaAs distributed Bragg reflectors (25 pairs below and 11 above), enabling efficient collection of ground-state emission from single QDs (Supplementary Fig.~\ref{Sample_structure}).

The QD studied here exhibited biexciton and exciton lifetimes of $\tau_{XX}=211\pm 5$\, ps and $\tau_{X}=405\pm 3$\, ps, respectively, and an exciton precession period of $T_{p}=470\pm 2$ ps, corresponding to a fine-structure splitting of $\Delta_{\mathrm{FSS}}=8.80\pm 0.04~\mu$eV.

\subsection*{Experimental setup}
The sample was mounted in a closed-cycle cryostat stabilized at $5\,\text{K}$. Excitation and collection were performed through a 60$\times$ microscope objective (NA = 0.85). Picosecond excitation pulses ($12\,\text{ps}$ duration, ${\sim}150\,\mu\text{eV}$ bandwidth) were generated by a synchronously pumped dye laser operating at $76\,\text{MHz}$. The laser wavelength was tuned to the biexciton two-photon excitation (TPE) resonance at $967.37\,\text{nm}$. A weak continuous-wave laser at $939.27\,\text{nm}$ optically depleted long-lived dark excitons~\cite{Schmidgall2015, Winik2017}.

Photoluminescence (PL) was spectrally filtered from the excitation laser and directed through a delay line that introduced a differential delay ~$\Delta t$ between the $XX^{0}$ and $X^{0}$ photons. The two paths were recombined and coupled into a fiber-based Mach–Zehnder interferometer (MZI). A polarizing beam splitter (PBS) separated horizontal ($H$) and vertical ($V$) polarizations into distinct arms, with a half-wave plate used to align the polarization basis.

An electro-optic phase modulator (EOM; $V_\pi\approx 4\,\text{V}$, $20\,\text{GHz}$ bandwidth) was placed in the $V$ arm. Upon recombination at the interferometer output, the differential phase produced a rotation about the $H$–$V$ axis. The EOM was driven by an arbitrary waveform generator (AWG) synchronized to the laser repetition rate, with residual timing jitter below $10\,\text{ps}$. The programmed waveform generated linear phase ramps of slope $\dot{\phi}\simeq\omega_X$, matching the exciton precession frequency and thereby cancelling the accumulated phase for both photons in the cascade.

\begin{figure}[h!]
\centering
\includegraphics[width=\textwidth]{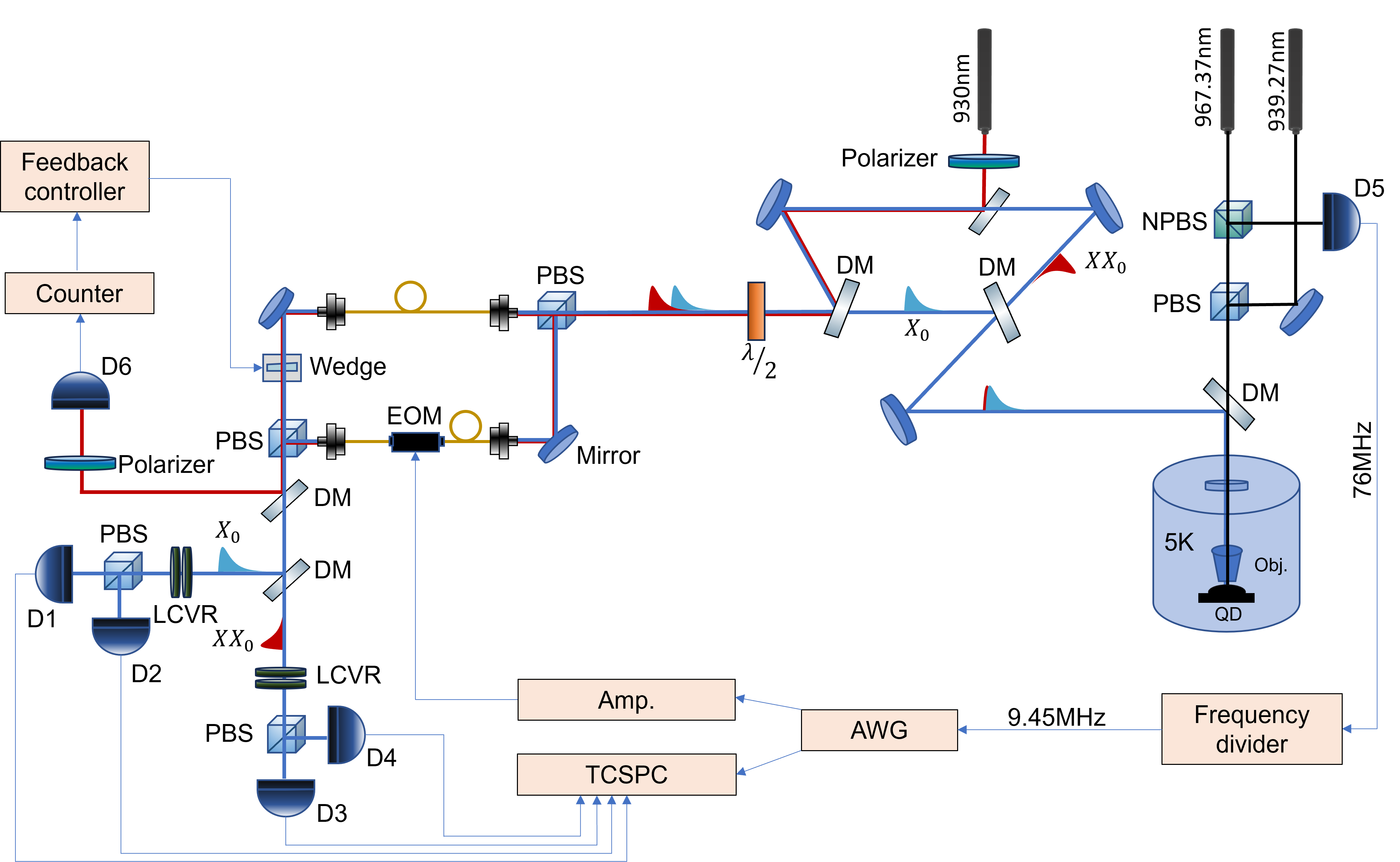}
\captionsetup{width=\textwidth}
\caption{
Schematic description of the experimental setup. NPBS, non-polarizing beam splitter; PBS, polarizing beam splitter; DM, dichroic mirror; TCSPC, time-correlated single-photon counter; obj., microscope objective; LCVR, liquid crystal variable retarder; $\lambda/2$, half-wave plate; Dn, detector n; CW, continuous wave; AWG, arbitrary waveform generator; EOM, electro-optic modulator; Amp., amplifier.
}
\label{full_experimental_schematics}
\end{figure}

Long-term phase stability of the interferometer was maintained using a $D$-polarized continuous-wave laser at $930\,\text{nm}$. The stabilization beam was injected before the MZI and extracted after it via long-pass and short-pass filters. The output was projected onto the anti-diagonal polarization, and a $4^\circ$ fused-silica wedge mounted on a motorized translation stage provided feedback control. A feedback loop maintained operation at a fringe minimum, achieving residual phase fluctuations of ${\sim}0.05$ rad over several hours (Supplementary Methods, Interferometer stability).

After the MZI, a narrow band-pass filter transmitted the biexciton photon while reflecting the exciton photon, allowing independent polarization analysis. Each photon passed through a pair of liquid-crystal variable retarders (LCVRs) followed by a polarizing beam splitter. Superconducting nanowire single-photon detectors (SNSPDs) recorded photon arrival times using a HydraHarp 400$^\text{TM}$ time-tagging module (Supplementary Fig.~\ref{full_experimental_schematics}). Background contributions were $<0.1\%$ of coincidence counts and were not subtracted.

\subsection*{Interferometer stability}
\label{SM:stability}
The Mach–Zehnder interferometer (MZI) used for dynamic phase modulation was actively stabilized using a
$D$-polarized continuous-wave laser at \SI{930}{\nano\metre}.
The stabilization beam was injected via a long-pass filter before the MZI and collected at the output through a short-pass filter,
then projected onto the anti-diagonal polarization.
A \SI{4}{\degree} fused-silica wedge mounted on a translation stage in one arm provided the feedback actuator.
The transmitted intensity was monitored and fed to a control loop that continuously adjusted the stage position
to maintain operation at a fringe minimum, thereby compensating for slow path-length drifts.

\begin{figure}[h!]
\centering
\includegraphics[width=0.6\textwidth]{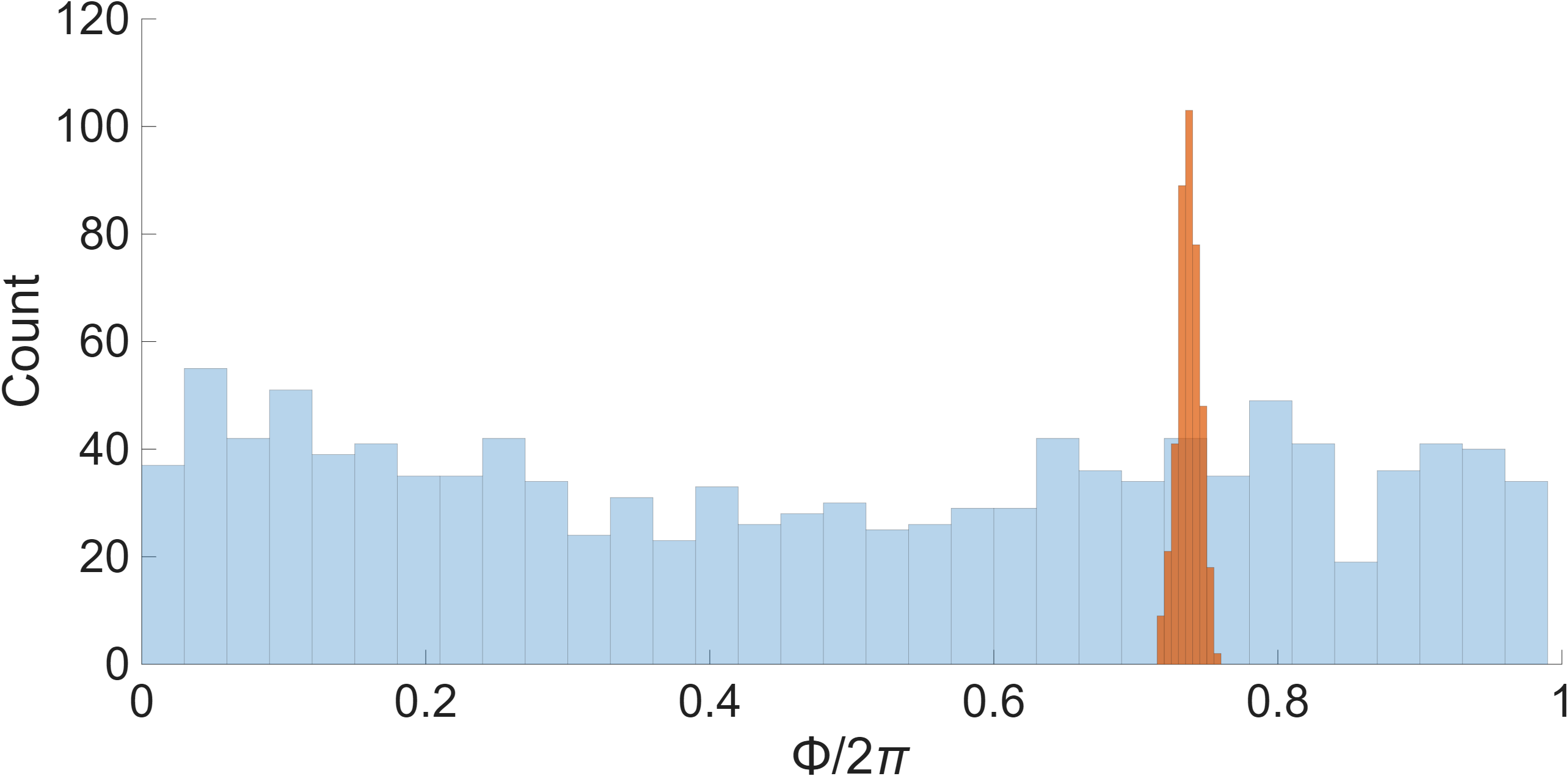}
\caption{\textbf{Interferometer phase stability.}
Histogram of the measured interferometer phase over several hours with active stabilization (orange) and without stabilization (blue).
Active feedback maintained operation near a fringe minimum with a phase standard deviation of \SI{0.05}{\radian}.}
\label{Supp_Stabilization_angle}
\end{figure}

The feedback bandwidth was approximately \SI{3}{\hertz}, sufficient to suppress thermal and mechanical fluctuations on timescales of minutes. Phase stability was monitored over several hours by recording the interference signal projected onto the $R$, $L$, $D$, and $A$ bases. With active stabilization, the interferometer operated at a fringe minimum with a phase standard deviation of \SI{0.05}{\radian}; without feedback, the phase drifted by several radians within minutes (Supplementary Fig.~\ref{Supp_Stabilization_angle}).

The stabilization scheme ensured phase-locked operation of the MZI throughout the long integration times required for polarization-state tomography.
The performance parameters reported here are identical to those used in the experiments presented in the main text.

\subsection*{Density matrix tomography}
Quantum-state tomography of the $XX^{0}$–$X^{0}$ photon pair was performed using 36 polarization projection settings spanning the $\{H,V,D,A,R,L\}$ bases for both photons~\cite{Kwiat2001}. Coincidence histograms were accumulated as a function of the photon arrival-time difference. A maximum-likelihood algorithm reconstructed a physical two-photon density matrix for each selected temporal detection window $t_W$. Statistical uncertainties were estimated assuming Poissonian count statistics and propagated through the reconstruction. Systematic uncertainties associated with polarization-setting inaccuracies were incorporated using a conservative $\pm3^\circ$ angular tolerance.

Negativity $\mathcal{N}$ was evaluated for each reconstructed density matrix using the Peres–Horodecki partial-transpose criterion~\cite{Peres1996,Vidal2002}. Equal temporal windows were used for $XX^{0}$ and $X^{0}$ photons to characterize the dependence of entanglement on the integration time.

\begin{figure}[h!]
\centering
\includegraphics[width=\textwidth]{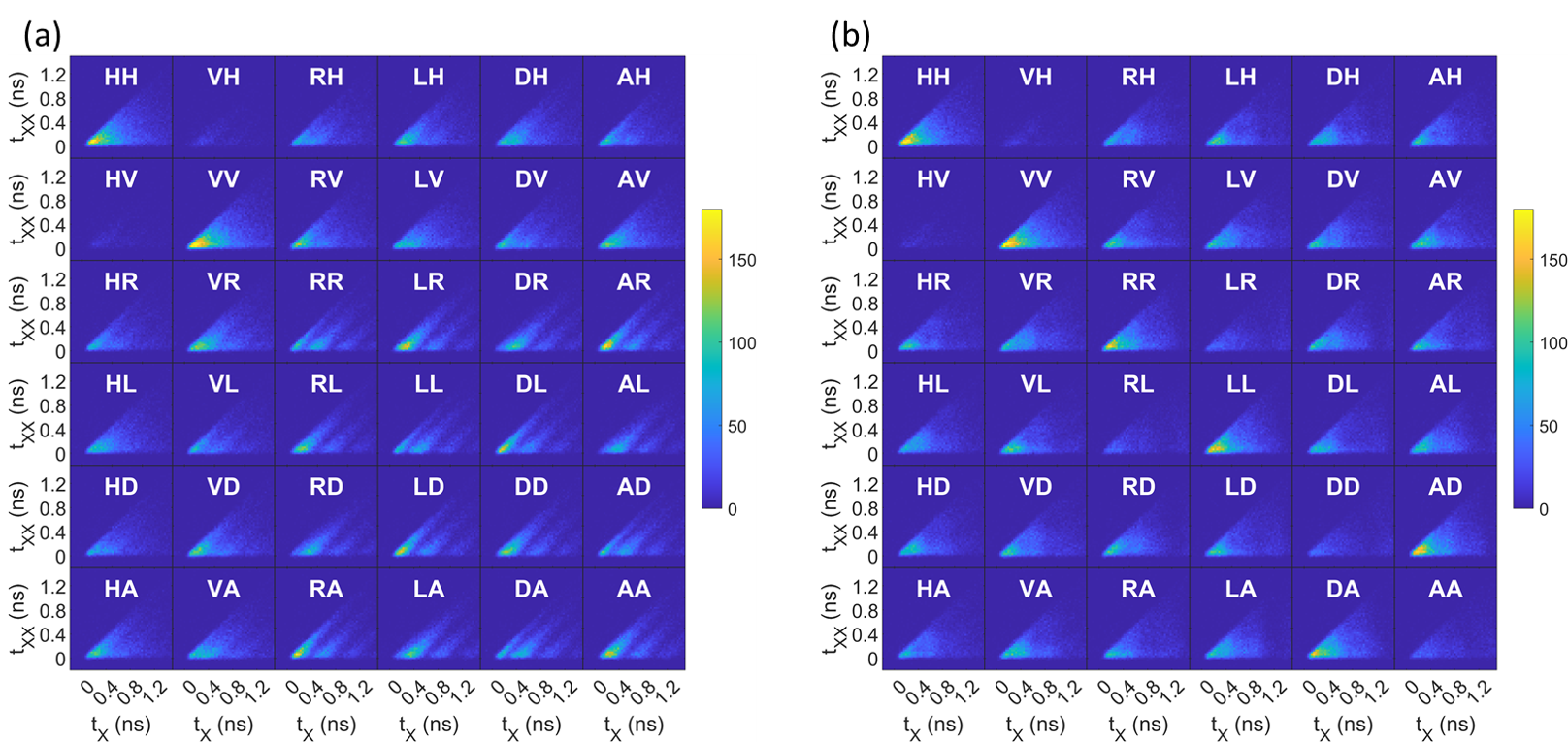}
\captionsetup{width=\textwidth}
\caption{
    Full set of coincidence measurements of biexciton-exciton photon pairs.
    (a) Coincidence measurements without DPM, and (b) coincidence measurements with DPM. Polarization settings are labeled as $P_{XX}P_X$ (e.g., VV, LR), where the first letter denotes the biexciton photon and the second the exciton photon. The color scale indicates the number of coincidence counts.
}
\label{Methods_Maps_DPM}
\end{figure}

Supplementary Fig.~\ref{Methods_Maps_DPM} presents the complete set of coincidence measurements of biexciton-exciton photon pairs used for the state tomography.

\section*{Supplementary Note 1: Dynamic phase modulation model}

We outline the theoretical framework for dynamic phase modulation (DPM) applied to the biexciton–exciton $\big( XX^0 - X^0\big )$ cascade in a semiconductor quantum dot.

\subsection*{Single-exciton phase evolution}

A resonant $\pi$-pulse with polarization $P = \alpha H + \beta V$ $\big (|\alpha|^2 + |\beta|^2 = 1\big )$ excites the exciton into
\begin{equation}
    \ket{\psi_X(0)} = \alpha \ket{X_H} + \beta \ket{X_V}.
\end{equation}
Here \(\ket{X_H}\) and \(\ket{X_V}\) denote eigenstates split by the fine-structure splitting (FSS) \(\Delta = E_X^V - E_X^H\).
During its radiative lifetime \(\tau_X\), the exciton precesses with angular frequency \(\omega_X = \Delta / \hbar\):
\begin{equation}
    \ket{\psi_X(t)} = \alpha \ket{X_H} + e^{-i\omega_X t}\beta \ket{X_V}.
\end{equation}
The emitted photon inherits this relative phase:
\begin{equation}
    \ket{\psi_{\mathrm{ph}}(t)} = \alpha \ket{H_X} + e^{-i\omega_X t_X}\beta \ket{V_X},
\end{equation}
where \(t_X\) is the photon emission time.
Applying a dynamic phase shift
\begin{equation}
    \Delta\phi_X(t) = \omega_X t_X
\end{equation}
between the \(H\) and \(V\) polarizations removes the time-dependent term, yielding a stationary photon state
\begin{equation}
    \ket{\psi_{\mathrm{ph}}(t)} = \alpha\ket{H_X} + \beta\ket{V_X}.
\end{equation}
Hence, DPM cancels the exciton precession when the applied phase slope satisfies \(\dot{\phi}_X = \omega_X\).

\subsection*{Biexciton–exciton cascade}

Excitation of the biexciton generates a sequential photon cascade:
\begin{equation}
    XX^0 \rightarrow X^0, \qquad
    X^0 \rightarrow 0.
\end{equation}
Immediately after the biexciton photon emission at time \(t_{XX}\), the intermediate photon–exciton state is
\begin{equation}
    \ket{\psi_{ph-X}(t)} \propto
    \ket{H_{XX}X_H} +
    e^{-i\omega_X (t - t_{XX})}\ket{V_{XX}X_V}.
\end{equation}
After the exciton decays at time \(t_X\), the resulting two-photon state is
\begin{equation}
    \ket{\psi_{2ph}(t)} \propto
    \ket{H_{XX}H_X} +
    e^{-i\omega_X (t_X - t_{XX})}\ket{V_{XX}V_X}.
\end{equation}
Because both $t_{XX}$ and $t_X$ are stochastic, dynamic phase modulation must be applied independently to each photon.

\subsection*{Dynamic phase compensation}

To restore entanglement, complementary DPMs are applied to the two photons:
\begin{equation}
    \Delta\phi_{XX}(t) = -\omega_X t_{XX}, \qquad
    \Delta\phi_X(t) = +\omega_X t_X.
\end{equation}
These corrections cancel the random phase factor \(e^{-i\omega_X (t_X - t_{XX})}\), resulting in a stationary Bell state:
\begin{equation}
    \ket{\psi_{2ph}} =
    \frac{1}{\sqrt{2}}\big(
    \ket{H_{XX}H_X} + \ket{V_{XX}V_X}
    \big).
\end{equation}
In principle, the scheme removes the stochastic timing dependence and can operate for arbitrary FSS provided the applied phase ramp matches $\omega_X$ over the relevant temporal window.

Residual phase mismatch or nonlinear modulation slightly reduces coherence but does not alter the underlying principle.
This treatment follows the theoretical proposals of Varo \textit{et al.} and Fognini \textit{et al.}~\cite{Varo2022, Fognini18}.

\bibliographystyle{naturemag}
\renewcommand{\refname}{Supplementary References}
\bibliography{refs}